\documentclass[journal=jacsat,manuscript=article]{achemso}
\usepackage{graphicx,color}
\usepackage{dcolumn}
\usepackage{float}

\usepackage{placeins}
\usepackage{wrapfig, blindtext}
\usepackage[dvipsnames]{xcolor}
\usepackage{xcolor}
\usepackage{booktabs,siunitx}
\usepackage{multirow,tabularx,booktabs} 
\usepackage[titletoc,toc,title]{appendix}
\usepackage{pst-node,lipsum}
\usepackage[caption=false]{subfig}
\usepackage[font=small,labelfont=bf,tableposition=top]{caption}
\usepackage[compact]{titlesec}         
\titlespacing{\section}{0pt}{0pt}{0pt} 
\AtBeginDocument{
	\setlength\abovedisplayskip{0pt}
	\setlength\belowdisplayskip{0pt}}

	\title{Revealing the role of van der Waals interactions in thiophene adsorption on copper surfaces }
	
	\author{Abhirup Patra}
	\affiliation{Department of Physics, Temple University, Philadelphia, PA 19122}
	\alsoaffiliation{School of Materials Science and Engineering, Georgia Institute of Technology, Atlanta, GA 30308}
	\email{abhirup.patra@hotmail.com}

	\author{Adrienn Ruzsinszky}
	\affiliation{Department of Physics, Temple University, Philadelphia, PA 19122}

\begin{document}
			
	\begin{abstract}
     Accurate modeling of electronic and structural properties of organic molecule-metal interfaces are challenging problems because of the complicated electronic distribution of molecule and screening of charges at the metallic surface. This is also the reason why the organic/inorganic system can be engineered for several applications by fine-tuning the metallic work function. Here, we use density-functional theory (DFT) calculations with different level of functional approximations for a systematic study of thiophene interacting with Copper surfaces. In particular, we considered adsorbed structures with the thiophene molecule seated on the top site, with the S atom of the molecule located on the top of a Cu atom. In this work, we find that the weak chemisorption hypothesis of thiophene binding on the copper surface is well justified by the two meta-GGAs-based approximations, SCAN and SCAN+rVV10. PBE-GGA and TM meta-GGA describe it as a physisorption phenomenon by significantly underestimating the adsorption energies. Calculated adsorption energy curves reveal that non-local dispersion interaction between the molecule and metallic surface predominantly controls the bonding mechanism and thus, modifies the copper's work function. Our results imply that semi-local functionals without any kind of van der Waals (vdW) correction can often misinterpret this as physisorption, while, a fortuitous error cancellation can give a right description of this adsorption picture for a wrong reason as in the case of SCAN. The calculated density of states of the adsorbed molecule shows that the long-range vdW correction of SCAN+rVV10 causes more than enough hybridization between the \textit{p} orbitals of S atom and the copper \textit{d}-bands and therefore overestimates the adsorption energies by an average of 16\%.
    
	\end{abstract}

	\maketitle
	\section{\label{sec:intro}INTRODUCTION}
	\vspace*{0.25cm}
     Theoretical modeling of the molecule-metal system is a fundamental topic of research in nano-electronics \cite{thiophenemoleculardevice1,thiophenemoleculedevice2}, and heterogeneous catalysis. Bonding of different organic molecules to surfaces gives useful insight into many mechanisms that are fundamentally important. In this context, adsorption of thiophene (C\textsubscript{4}H\textsubscript{4}S) on different metallic surfaces has been studied within both theoretical and experimental frameworks. While the strength of chemical interactions and the extent of charge transfer between the thiophene and metal are important features for designing new nano-electronic devices or desulfurization processes, an important catalytic reaction used for purification of natural petroleum takes place via C-S bond breakage \cite{desulphur-1}. Thiophene adsorbed on transition metallic surfaces can be used as a prototype for detailed investigation of such desulfurization mechanism of petroleum products. In this work, we will focus on issues and features of thiophene adsorbed on the three different crystalline faces of copper within the density functional theory (DFT) framework.

	Adsorption of thiophene on a Cu surface greatly depends on how electronic distribution changes due to the binding of S atom of thiophene with the surface Cu atom. From the chemical point of view, S atom of the thiophene molecule can bind with the metallic system either by donating the lone pair of electrons from the sulfur atom to the metal surface or by delocalizing the $\pi$ type electron of the thiophene ring. Experimentally, the thiophene adsorption \cite{thiopheneNIXSW-NEXFAS-1,thiopheneTPD} has been studied on Cu(111) surfaces using normal incidence x-ray standing wave (NIXSW) \cite{thiopheneNIXSW-NEXFAS-1}, near-edge x-ray absorption fine structure (NEXAFS), and temperature-programmed desorption (TPD) \cite{thiopheneTPD} measurements. These works have revealed different structural information of the adsorbed molecule on the metallic surface, such as molecule-surface distance, tilt angle, the adsorption site. Like many other organic molecules, adsorption of the thiophene on a noble metal is often considered as a pure physisorption type. While the above mentioned experimental works cannot give us any clear information about the structural configuration of the adsorbed thiophene molecule, the binding nature of thiophene with the metallic surface is confirmed to be a weak chemisorption type. Apart from this, several theoretical studies of single thiophene adsorbed on the metal surface suggest that strong chemisorption with desulfurization can be seen in the case of Ni (100) \cite{thiopheneNiCuPd100DFT,thiopheneNi110DFT} surface.
    Many recent theoretical works \cite{erinjohnsonthiophene, Hu2014} suggest that the interaction between thiophene and the metallic surface is mostly dispersion mediated, and dispersion corrected functionals must be used which correctly accounts for the van der Waals (vdW) interaction. Therefore, standard exchange-correlation approximations without vdW correction can be severely wrong when predicting the nature of molecule-to-metal bonding. For example, the generalized gradient approximation (GGA) like Perdew-Burke-Ernzerhof (PBE) performs badly for typical vdW systems where long-range vdW interaction is significant. Hence, they often fail to predict correct binding example giving too small adsorption energies. The nature of vdW interactions between the thiophene molecule and copper surface is non-local in nature. Therefore, using a vdW corrected functional that truly captures this behavior of the molecule-metal interaction, yields more accurate adsorption energies compared to the regular Local density approximation (LDA) and GGAs. The vdW-DF functional family \cite{vdWdf1} seamlessly captures the asymptotic region of this long-range vdW interaction utilizing a kernel that connects two interacting densities via a non-local integration \cite{thiopheneCu110PRLsony}. In a few situations, the weak vdW interaction between an organic molecule and a metal surface is also well described by more popular and less expensive but semi-empirical pairwise correction methods such as DFT-D functionals \citep{Grimmelongrange,grimme2011effect, DFTD3}.
	These functionals show significant improvement in describing adsorption energies of thiophene adsorbed on Cu (111) surface \citep{thiophene111surfacegrob,callsenthiophene}. However, it is important to note that these studies are only limited to GGA functionals. Therefore, it is necessary to explore higher-level approximations of density functionals on Perdew's Jacob's ladder \cite{jacobsladder}. In this work, we explored the performance of some newly developed meta-GGA functionals like strongly constrained and appropriately normed (SCAN) \cite{SCAN}, Tao-Mo \cite{Tao-Mo} and a vdW-corrected form of SCAN, SCAN+rVV10  \cite{rVV10}. Furthermore, we will show that SCAN can predict adsorption energy and adsorbed geometry in a very good agreement with the available experimental results. Specifically, we investigated the adsorption and desulfurization mechanism of a thiophene molecule on three different low-index copper surfaces.  
	
    This paper is organized as follows. In the ``Computational Details'' section we describe the computational framework utilized in this work and provide details of the geometry used in this work. In ``Results \& Discussions'', we benchmark the performance of functionals for the Cu (111), Cu (110), and Cu (100) surfaces. Finally, in the ``Conclusion'' section we summarize the insights gained in this work and provide a perspective of how advanced density functionals can be used to understand complicated interactions at the interface of a metal and molecule.
	
	\vspace*{0.25cm}
	\section{COMPUTATIONAL DETAILS}
	\vspace*{0.25cm}

\begin{figure*}[!ht]
		\centering
		\includegraphics[width=1\textwidth]{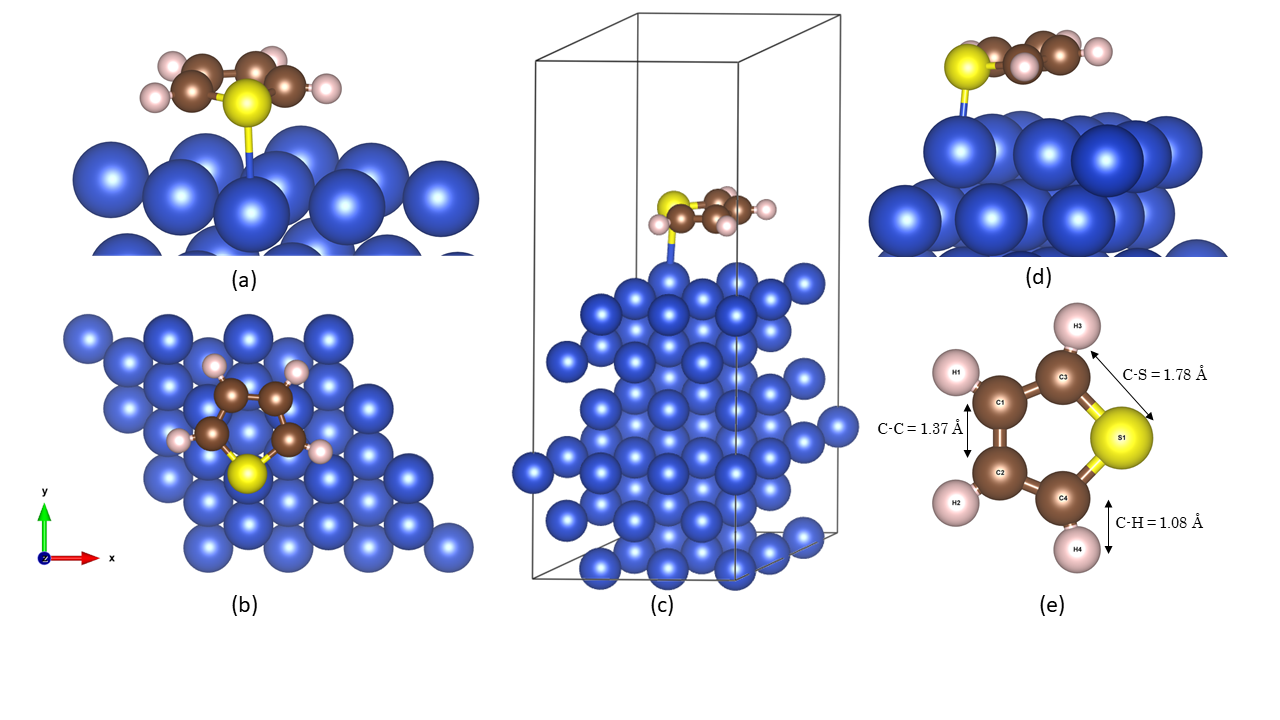}

		\caption{This figure shows the molecule-surface model for the thiophene/Cu (110) case. In figure (a), the surface slab with 6 layers and a thiophene molecule sitting at the top is shown. Figures (b) and (c) show the top and side view of the same system. In figure (c) 2.60 $\SI{}{\angstrom}$ is the Cu-S distance used to place the thiophene on Cu (110) surface.}
		\label{fig:geo}
		
	\end{figure*}

    To model the molecule-metal system we have considered copper slabs of six atomic layer thickness and an adsorbed single thiophene molecule such that the S atom of thiophene sits at the top of a surface copper atom. Although other adsorption configurations are still relevant from the practical point of view, we want to emphasize that the top configuration is reported as the favorable site by many recent studies \cite{callsenthiophene}. The surface slabs were built using a $4 \times 4$ surface unit cell, and 12 $\SI{}{\angstrom}$ of vacuum separating a slab from its periodic image as shown in Fig. \ref{fig:geo}. In that figure, we show a different orientation of the conjugated system. Specifically, Fig.\ref{fig:geo} shows the configurations of the thiophene/Cu(110) for the SCAN calculation. These metallic slabs are generated using theoretical lattice constants obtained from fitting the Birch-Murnaghan equation of state (BM-EOS) to the energy-volume data \citep{PatraPNAS2017, Tao-Mo}. Molecular coverage on metal surfaces often controls both the adsorption geometry and energy. However, it is important to note that very large molecular coverage increases the computational cost. Furthermore, a coverage dependent study \cite{milligan2001complete} shows that at low coverage the thiophene tends to align parallel to the copper surface, but, high-coverage introduces more covalent interaction resulting in more tilting of the adsorbed molecule. We have chosen a coverage of 0.03 ML for our study as we intend to gain useful information about the behavior of different functionals.

	The first-principles calculation done in this work is based on the plane wave density functional theory (PAW-DFT) framework. The Vienna Ab Initio Simulation Package (VASP) \cite{vasp1,vasp2,vasp3} (version 5.4.4) is used to perform all DFT calculations reported in this work. The electron-ion interaction is approximated using the pseudopotential projector augmented wave (PAW) formalism \cite{blochl1994projector} within the plane-wave implementation of Kohn-Sham \cite{KOHN-DFT1} scheme. The molecule-metal systems were optimized by relaxing the top three layers of the metallic slab and the molecule. A kinetic-energy cutoff of 450 eV is used for the wave functions, and the Brillouin zone is sampled using $\Gamma$ centered  $6 \times 6 \times 1$ k-points for the slab, and $1 \times 1 \times 1$ for the molecule in a box of size $13 \times 14 \times 16 $ $\SI{}{\angstrom}^{3}$. The optimization of total energies for each adsorption configuration is performed using the Conjugate Gradient (CG) algorithm with the force criterion on each atom set for the convergence to be 0.005 eV/$\SI{}{\angstrom}$. The adsorption energy of the thiophene molecule is calculated using--
	\begin{equation}
	E_{ads} = E_{mol/metal}- (E_{metal} + E_{mol}),
	\label{eq:organic-1}
	\end{equation}
	 where, the $E_{mol/metal}$ is the total energy of the molecule-metal system, $E_{metal}$ is the total energy of clean surface slab, and, $E_{mol}$
	is the total energy of the molecule in the gas phase respectively.

	\vspace*{0.25cm}
	\section{\large{RESULTS \& DISCUSSIONS}}
		\vspace*{0.25cm}
		\FloatBarrier
	\subsection{Adsorption energies and configurations}
	\label{sec:adsorption}
\begin{figure*}[!h]
   	
   	\centering
   	\includegraphics[width=1.00\textwidth]{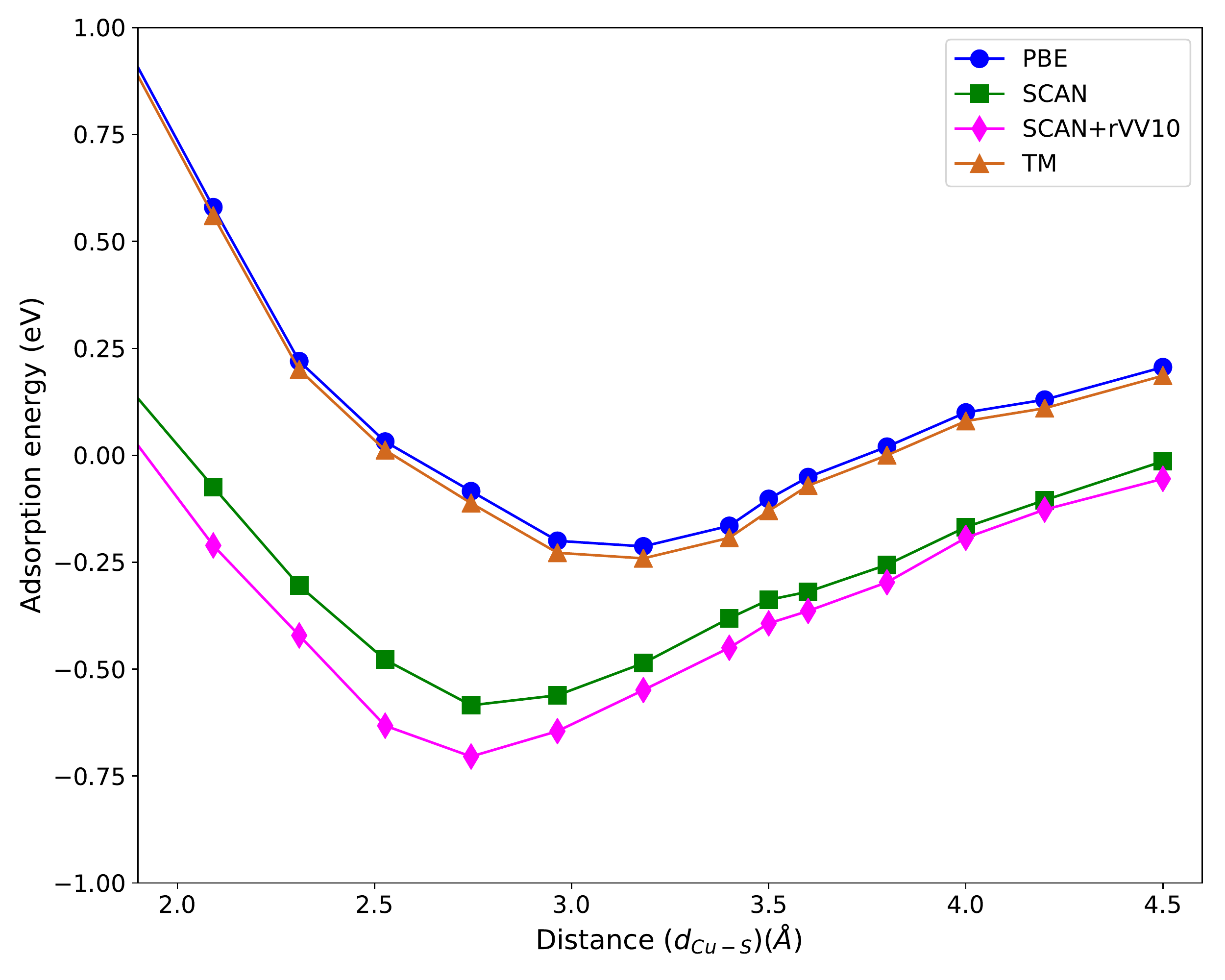}
   	\caption{The adsorption energy curve of thiophene bonded with the Cu(111) surface is plotted with respect to the Cu-S distance for different functionals. }
   	\label{fig:binding-111}
   	
   \end{figure*}

\begin{figure*}[!ht]
   	\centering
   	\includegraphics[width=1\textwidth]{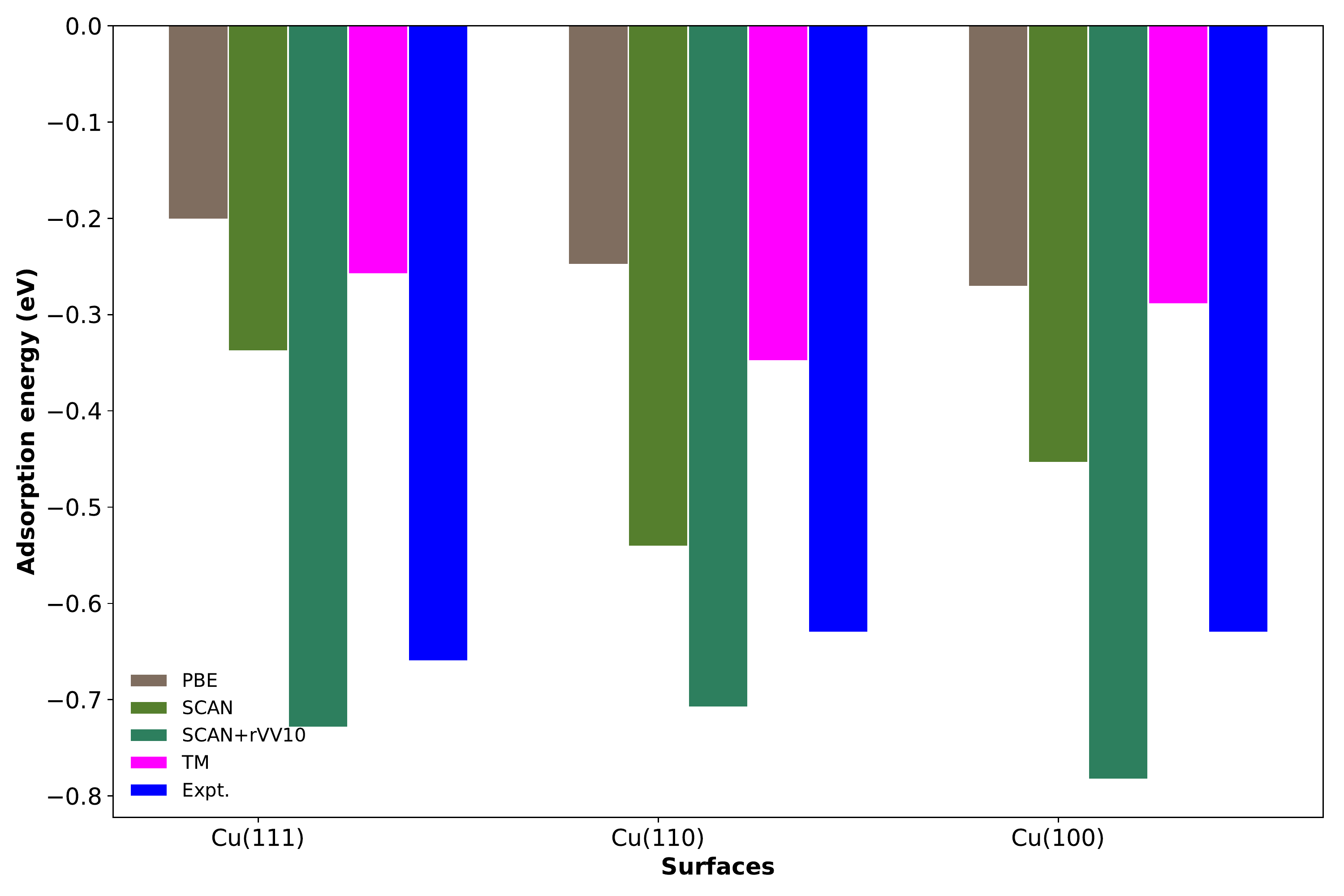} 
   	\caption{The adsorption energy of Thiophene on Copper surfaces as tabulated in Table \ref{tab:organic}. PBE systematically underestimates the molecule-metal weak interactions and hence predicts too weak Cu-S bonding. Intermediate-range vdW interaction present in SCAN capture this weak interaction but too much of long-range correlation of SCAN+rVV10 overestimates the binding. 
   	}
   	\label{fig:ads-bar}
   \end{figure*}

       First, we investigate the stable adsorption site for a single thiophene molecule on the Copper surface. After considering the top, hollow and bridge position we find that the configuration where the S atom of the molecule sits on top of a Cu atom is the most stable position for the low-coverage limit. A similar finding is also reported in a relevant work by Callsen et al.\cite{callsenthiophene} using a PBE calculation. In Table.\ref{tab:organic}, calculated adsorption energies ($E_{ads}$) are shown. A bar plot showing the variation of the predicted adsorption energies is shown in Fig.\ref{fig:ads-bar}. From the tabulated values of $E_{ads}$, it is clear that PBE largely underestimates the ``attractive'' van der Waals interaction between the molecule and the surface. As a result, the PBE adsorption energies are largely underestimated for the Cu(111) surface. We find that the calculated $E_{ads}$ value for Cu(111) surface using PBE in this work is higher than the previously reported values \cite{Hu2014,callsenthiophene} by 0.1 eV. The stable binding distances predicted by PBE are much larger compared to the experimentally observed binding distance (See Table.\ref{tab:organic}). PBE puts the thiophene at a height 0.2 $\SI{}{\angstrom}$ higher than the reference value, which completely describes the underbinding of thiophene with Cu(111) surface. This proves the fact that the lack of vdW interaction in PBE results in a poor description of weak but important dispersion interaction in sparse matters like hydrocarbons \cite{Liu_2013}. However, the tilting angle of the adsorbed thiophene molecule is well described by PBE as can be seen in Table.\ref{tab:organic}. While meta-GGA functionals are more complicated compared to PBE and designed to handle this kind of complicated situation, they still belong to the semi-local class of the density functional approximation. Being one of the most sophisticated and non-empirical meta-GGA functional, SCAN has shown remarkably good performance for the cases \cite{SR16} where many other members of the same class of density functionals perform poorly. We find that SCAN can accurately capture the weak vdW interactions between metallic Cu $d$ orbitals and de-localized $\pi$ orbitals of thiophene aromatic ring. For the Cu (111) surface SCAN gives adsorption energy of -0.538 eV, underbinding experimental value by only 0.13 eV but more accurate compared to PBE. SCAN predicts the equilibrium binding distance of the Cu-S 0.04 $\SI{}{\angstrom}$ smaller than the experimentally measured distance of 2.62 $\SI{}{\angstrom}$ and the tilt angle of 20.25\textsuperscript{o}. Thiophene binding energy is overestimated by about 30\% in SCAN+rVV10 calculation. The attractive vdW captured by the non-local rVV10 kernel brings the thiophene molecule much closer to the metal surface and much less titled as can be seen from the adsorption energy curve of Fig.\ref{fig:binding-111}. Performance of the recently developed Tao-Mao meta-GGA functional is similar to the PBE functional, it underestimates the adsorption energy and overestimates the binding distance significantly for Cu(111) surface. Although, the TM functional has shown much better performance for clean surfaces compared to the PBE, for a conjugated molecule-metal system.

   \FloatBarrier
   \begin{table*}[!ht]	
   	\begin{tabular}{|c|c|c|c|c|c|c|c|c|c|}
   		\hline
   		\multirow{3}{*}{Methods} & \multicolumn{3}{c|}{Cu(111)} & \multicolumn{3}{c|}{Cu(110)} & \multicolumn{3}{c|}{Cu(100)}  \\
   		\cline{2-10}
   		& $E_{ads} $ & $d_{Cu-S} $  & $\delta^{o}$ &  $E_{ads} $ & $d_{Cu-S} $  & $\delta^{o}$ & $E_{ads} $ & $d_{Cu-S} $  & $\delta^{o}$ \\
   		& $(eV)$ & $(\SI{}{\angstrom})$  &  & $(eV)$ & $(\SI{}{\angstrom})$  &  & $(eV)$ & $(\SI{}{\angstrom})$  &   \\
   		\hline
   		PBE &  -0.201 & 2.80 & 20.01   &  -0.248 & 2.75 & 0.5 & -0.271 & 2.47 & 1.47  \\
   		\hline
   		SCAN & -0.538  & 2.58 & 20.25 &  -0.541  &   2.56 & 1.05 & -0.454 &  2.46 & 1.41 \\
   		\hline
   		SCAN+rVV10 &  -0.729  & 2.51 & 22.88 &  -0.748  & 2.54 & 7.25   & -0.783 & 2.44 & 2.69 \\
   		\hline
   		TM &  -0.258 & 2.77 & 19.85 &  -0.348 & 2.75 & 1.47 & -0.289 &  2.32 & 0.55 \\
   		\hline
   		Expt. & \textbf{-0.66}\textsuperscript{\cite{thiopheneNIXSW-NEXFAS-1}}  &  2.62  & 26   & \textbf{-0.63} & & & \textbf{-0.63}\textsuperscript{\cite{thiopheneNIXSW-NEXFAS-1}}  & 2.42 & 0 \\
   		&   & $\pm$ 0.03 & $\pm$ 5 & \textbf{$\pm$ 10\%}\textsuperscript{\cite{thiopheneNIXSW-NEXFAS-1}} & & &  &  $\pm$ 0.02 &  $\pm$5  \\
   		\hline
   	\end{tabular}
   	\caption{Adsorption energies (in eV) of thiophene on Cu (111),  Cu (110) \& Cu (100) surfaces. Experimental adsorption energies are shown in bold number \cite{thiopheneNIXSW-NEXFAS-1}.}
   	\label{tab:organic}
   \end{table*}
   
   \FloatBarrier
   
\begin{figure*}[!h]
   	
   	\centering
   	\includegraphics[width=1.00\textwidth]{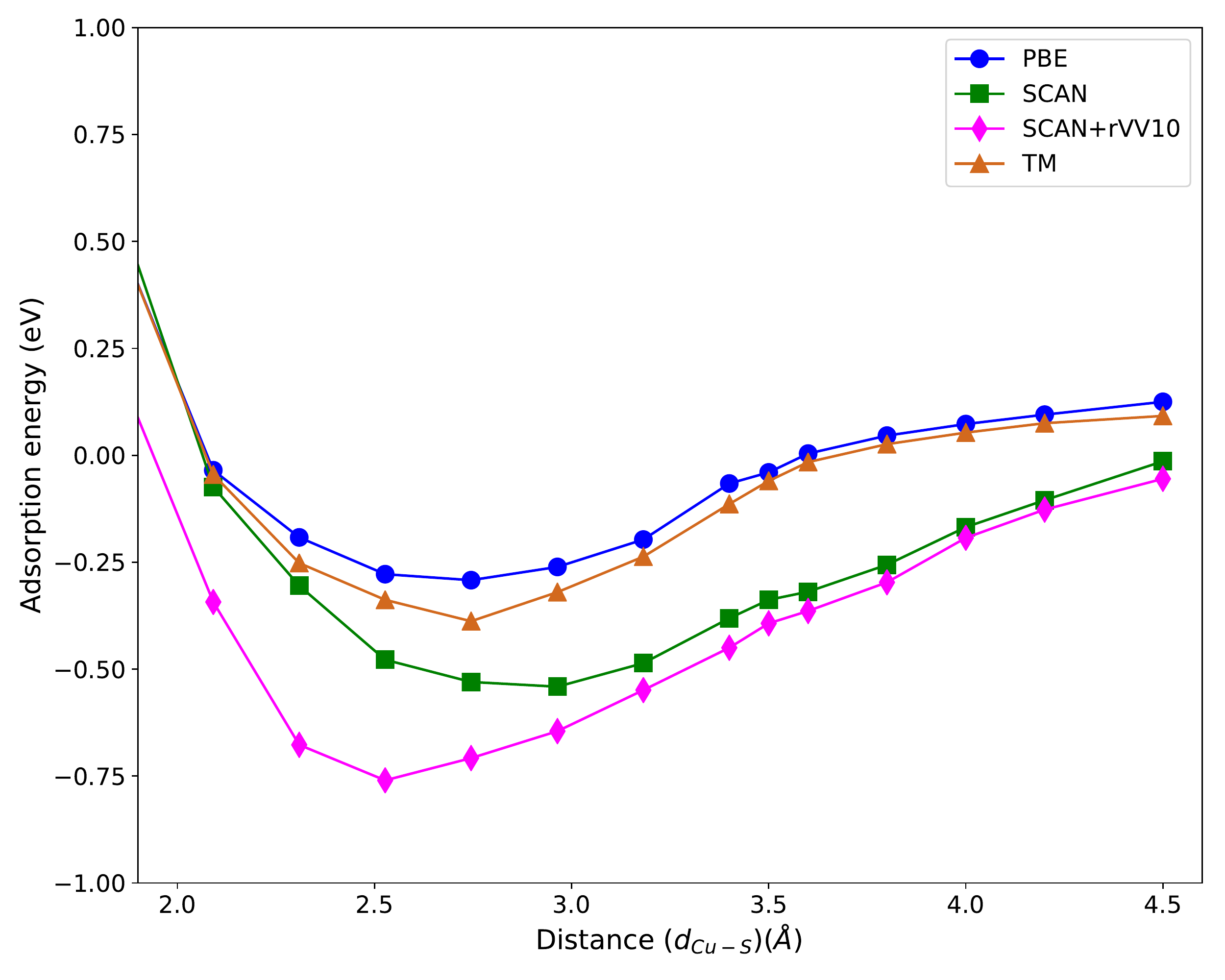}
   	
   	\caption{The adsorption energy curve of the thiophene bonded with the Cu(110) surface is plotted with respect to the Cu-S distance for different functionals. }
   	\label{fig:binding-110}
   	
   \end{figure*}

    Similar performance of PBE functional can be seen from Table.\ref{tab:organic} and the adsorption energy curve of Fig.\ref{fig:binding-110}. A systematic underbinding of the molecule to the metal surface is caused by larger binding distance (2.75 $\SI{}{\angstrom}$) predicted by PBE. SCAN binds thiophene more strongly to the surface. 
	 
	In a recent study, it has been suspected that the self-consistent SCAN density is accurate for a clean metallic surface but for a metal-molecule system like CO adsorbed on transition metal surface leads to spurious charge transfer from metal to molecule \cite{COMETALS}. We will further explore the role of the electronic structure of the adsorbed thiophene molecule in the next section. Nevertheless, comparing adsorption energies of SCAN with the ones from PBE calculation one can see that inclusion of more exact constraints made SCAN a better functional. Better performance of SCAN+rVV10 over PBE and SCAN can be seen clearly from Table.\ref{tab:organic}. For metallic surfaces, the asymptotic part of the attractive vdW potential at the surface can be accurately determined by the self-consistent SCAN+rVV10 density. The true many-body nature of the long-range electron correlation is approached within the self-consistent exchange-correlation potential of SCAN+rVV10 which significantly improves the work function by modifying the Fermi level of metal surface \cite{PNAS}. This is also true for a metal surface interacting with the organic molecule thiophene due to an expected error cancellation between the SCAN exchange and the correlation. While the reference value of $E_{ads}$ is not available for Cu(110), SCAN+rVV10 predicts much higher adsorption energy and shorter binding distance compared to PBE and SCAN. The Tao-Mao meta-GGA functional performance for Cu(110) surface is not very different from the Cu(111) one. 
   
   	\FloatBarrier

   \FloatBarrier

\begin{figure*}[!h]
	
	\centering
	\includegraphics[width=1.00\textwidth]{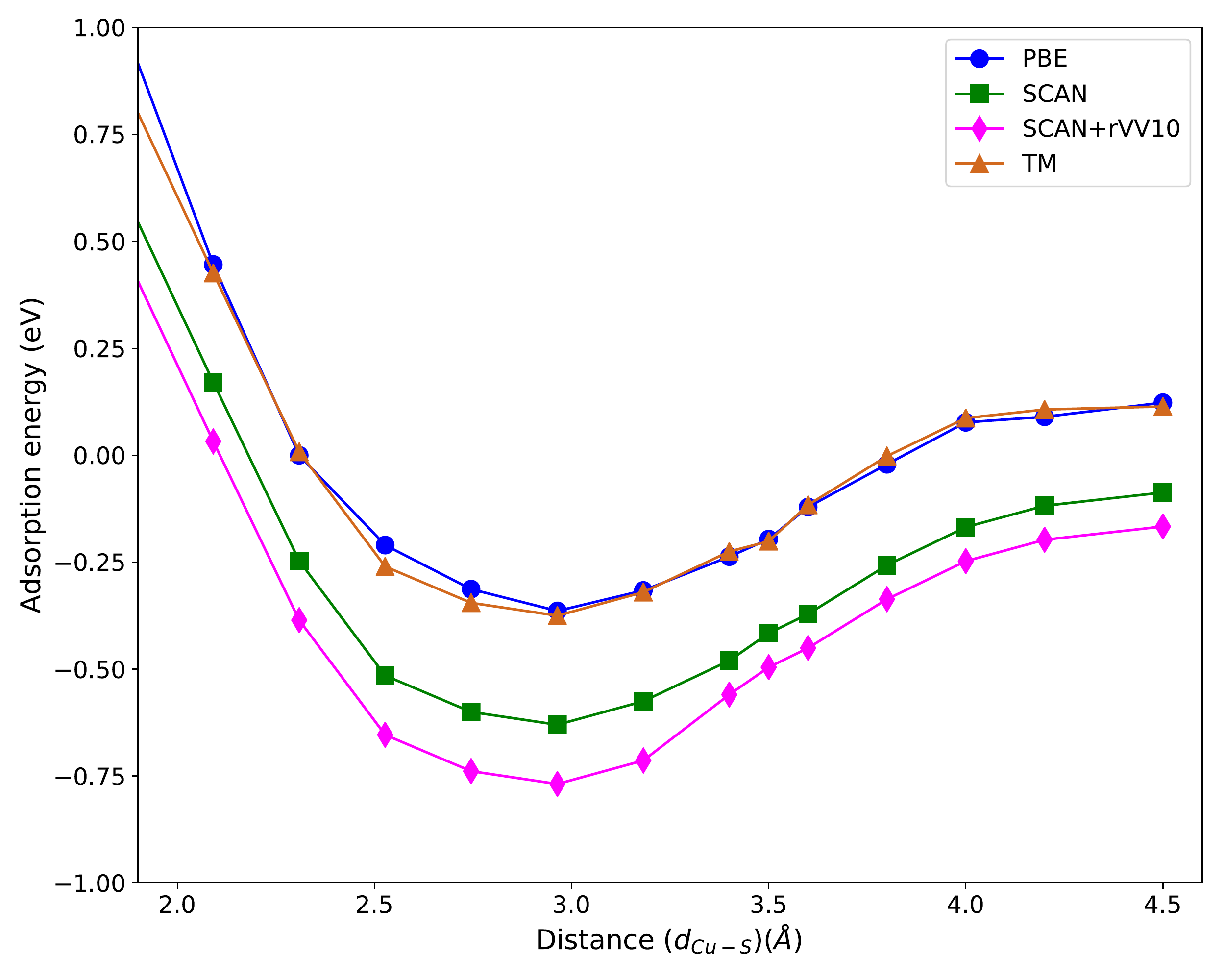}

	\caption{The adsorption energy curve of thiophene bonded with the Cu(100) surface is plotted with respect to the Cu-S distance for different functionals. }
	\label{fig:binding-100}
	
\end{figure*}

	The Cu(100) surface is not very different from the Cu(110) in terms of its work function. Therefore, we should also expect a similar performance of the functionals for this surface. PBE and TM underbind the reference adsorption energy of thiophene molecule by 42\% and 44\% respectively. A detailed discussion about the charge redistribution of the molecule-metal interface is given in \ref{sec:electronic}. Both PBE and TM keep the molecule at a much higher distance from the copper surface. Table.\ref{tab:organic} shows the equilibrium molecule-metal distance and the tilt angle for PBE. A systematic increment in adsorption energy from PBE to SCAN to SCAN+rVV10 can be seen from the same table. Adding non-local vdW correction via the rVV10 kernel, SCAN+rVV10 binds the thiophene molecule much stronger to Cu(110) surface than SCAN. This strong molecule-metal interaction also leads to much larger tilting angle for the molecule. The adsorption energy curve for thiophene on Cu(110) for all these fuctionals is depicted in Fig.\ref{fig:binding-110} and that on Cu(100) in Fig.\ref{fig:binding-100}. 
		
\FloatBarrier

	\FloatBarrier
\subsection{Electronic Structure of the adsorbed system}
\label{sec:electronic}
    Electronic structure of the metal-molecule interface is largely affected by the nature of the organic molecule. Adsorption of a strong electron acceptor or donor hydrocarbon often causes charge transfer between the lowest unoccupied molecular orbital (LUMO) and the metallic substrate. The main challenge in modeling molecule-metal interface within the single-particle Kohn-Sham (KS) density functional framework is self-interaction error of the semi-local exchange-correlation functionals, which can place the LUMO of molecules too low and the highest occupied orbital (HOMO) too high causing significant spurious charge transfer. The dipole that builds up due to this electron transfer modifies the metallic work function. In this section, we focus on the modification of the electronic structure of the metal by studying work function change and the change in position of the HOMO, LUMO orbitals using atom projected density of states (pDOS) of the adsorbed molecule. 

\begin{figure*}[!htb]
	
	\centering
	\includegraphics[width=1.00\textwidth]{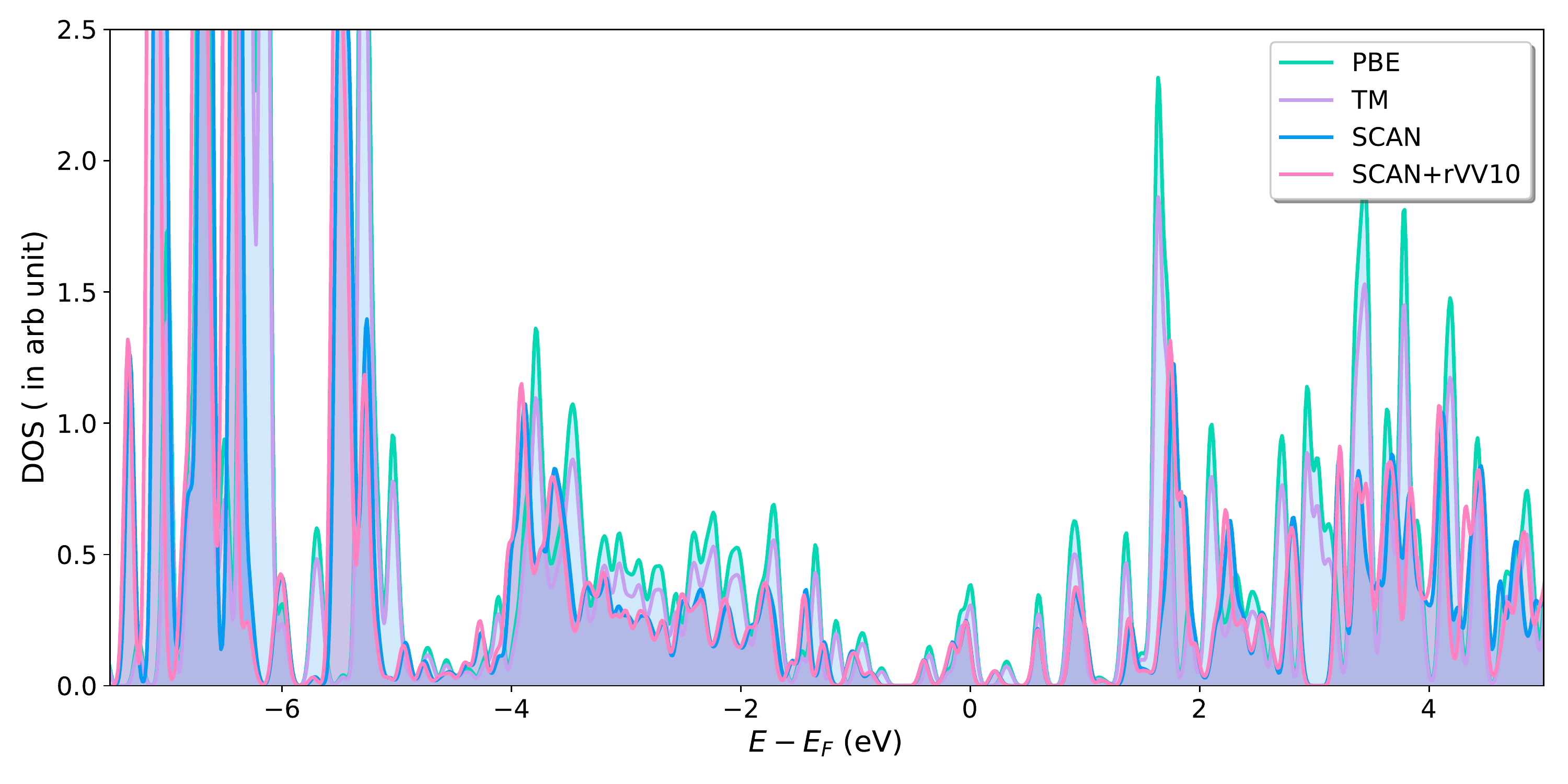}
	
	\caption{In this figure density of states of the adsorbed thiophene molecule are plotted. The adsorbed thiophene molecule is sitting on top of a copper atom of Cu(111) surface.}
	\label{fig:dos-111}
	
\end{figure*}

Figure~\ref{fig:dos-111} shows pDOS of the adsorbed thiophene molecule using PBE, TM, SCAN, and SCAN+rVV10. It is quite clear from these plots that PBE and TM allow the charge transfer from the molecule to the metal causing a shift in molecular HOMO closer to the Fermi level while both SCAN and SCAN+rVV10 make the electronegative S atom of thiophene pull more electrons from the metallic surface. This metal-to-molecule charge transfer causes a downward interface dipole moment leading to a decrement in metallic work function as can be seen from Table.\ref{tab:wf}. The work function decreased by 0.04 and 0.08 eV for PBE and TM but 0.07 and 0.11 eV for SCAN and SCAN+rVV10 respectively. This further facilitates the fact that weak but important vdW interactions between the molecule and metal surface can create a hybridization between the Cu d bands and the $\pi$ electrons of thiophene giving a vacuum level shift. 
  
\FloatBarrier
\begin{table*}[!ht]
	\begin{tabular}{l*{5}c}
		\toprule
		& PBE & SCAN & SCAN   & TM    \\
		&     &      & +rVV10 &  	    \\
		\midrule \midrule \addlinespace \\
		$\Delta \Phi_{111}$ & -0.04 &  -0.07	&  -0.11 	&   -0.08  \\
		$\Delta \Phi_{110}$ & -0.07 &  -0.10 	&  -0.15  	&  -0.04   \\
		$\Delta \Phi_{100}$ & -0.05 & -0.19		& -0.22		&   -0.12 	\\
		\hline
		\bottomrule
	\end{tabular}
	\caption{Change in work function ($\Delta \Phi =  \Phi_{Cu-Th}-\Phi_{Cu}$) with respect to clean surface of the conjugated system is tabulated here. The work function for both conjugated system ($ \Phi_{Cu-Th}$) and clean surface ($ \Phi_{Cu}$) are calculated with respect to the vacuum potential of the slab model. }
	\label{tab:wf}
\end{table*}

	Next, we will see that when thiophene adsorbed on Cu(110) surface there is a significant shift in SCAN+rVV10 DOS compared to the one predicted by PBE, TM and even SCAN. This shift in the molecular DOS indicates a strong electron-transfer from metal to the LUMO of thiophene. This further strengthens the fact that in the case of PBE and even SCAN the LUMO of the molecule is more than half-filled and molecular orbital energies are underestimated. PBE lowers the work function by 0.07 eV due to a different electronic arrangement at the interface compared to SCAN+rVV10. Surprisingly, TM doesn't alter the copper work function too much- work function of the adsorbed system is only 0.04 eV lower than the clean surface. Owing to the weak chemical nature of the adsorption in SCAN calculation a partial charge transfer between the Cu(110) surface and the molecule shifts Fermi level of the and hence lowers the work function by 0.10 eV. The vdW interactions in SCAN+rVV10 cause more charge transfer from filled Cu-d bands to the S of the molecule lowering Cu work function even more (see Table.\ref{tab:wf}). However, the change in work function also arises from the different description of metallic electronic structure by these methods. As previously shown \cite{PNAS}, screening at metallic surfaces is better described by SCAN and SCAN+rVV10 than PBE leading to a more accurate clean surface work function. Nevertheless, currently we do not have any reference experimental value for these changes in work function but a quantitative picture from these computational values can be inferred.

\FloatBarrier

\begin{figure*}[!h]
	
	\centering
	
	\includegraphics[width=1.00\textwidth]{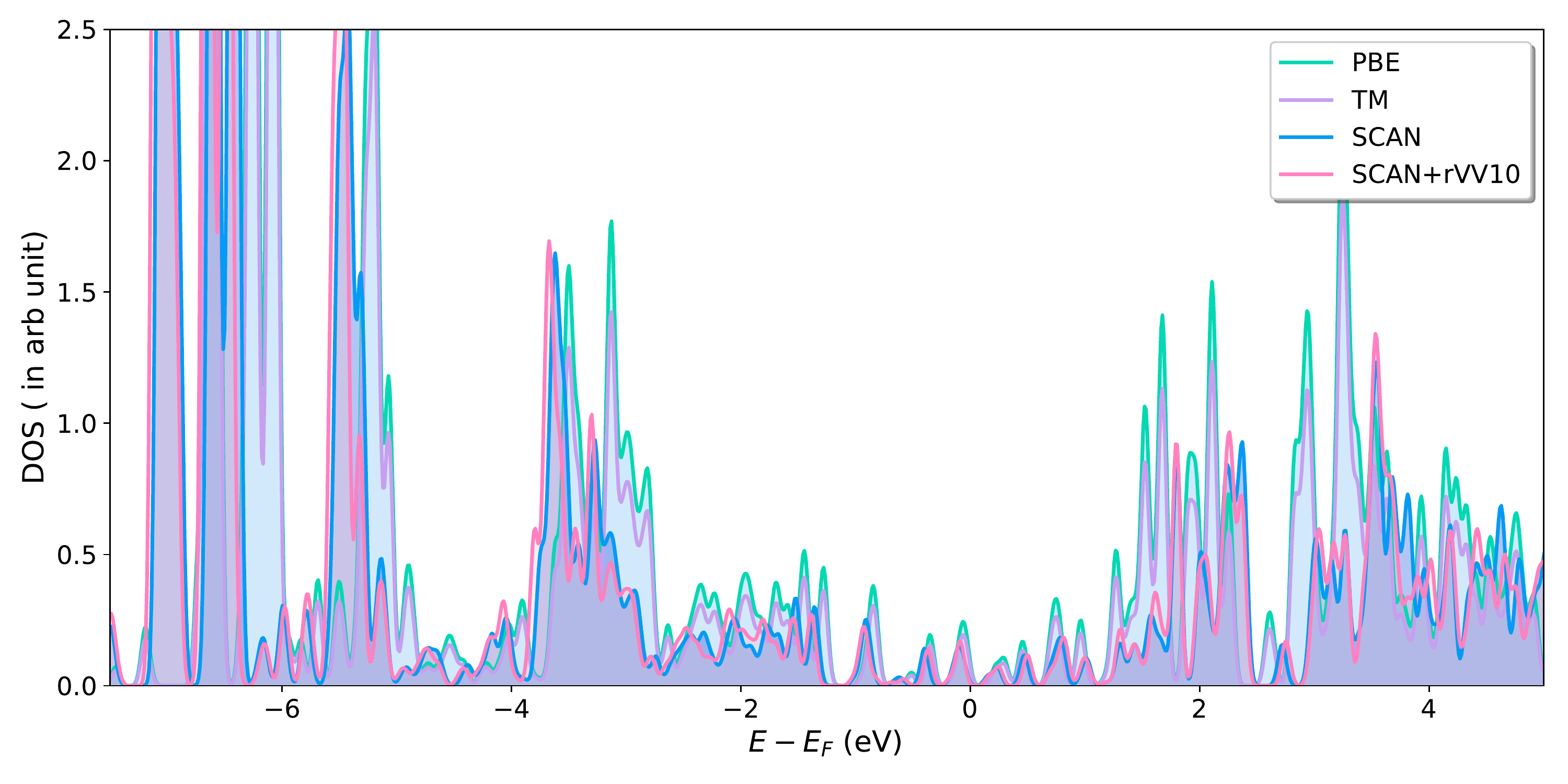}

	\caption{This figure captures density of states distribution of the thiophene molecule on Cu(100) surface. }
	
	\label{fig:dos-100}
	
\end{figure*}

    Figure \ref{fig:dos-100} shows the PDOS of adsorbed thiophene molecule on Cu(100) surface. Similar to the other two copper surface a strong hybridization between the Cu \textit{d}-states and the molecule is visible from this plot. The work function change ($\Delta \Phi$) for this surface is the highest for SCAN+rVV10 and lowest for PBE suggesting strong chemisorption of thiophene due to covalent bond formation between S and Cu (Please note that at present we do not understand the behavior of TM meta-GGA for this particular system). Although TM performs quite similar to PBE, putting the molecular orbitals at almost the same position with respect to the metallic surface. But, a greater change in work function and stronger adsorption energy by TM indicates that it can better describe the electron transfer compared to the PBE but not as good as SCAN or SCAN+rVV10. This can be fixed by adding of long-range vdW correction to the TM functional. 
\FloatBarrier
\begin{figure*}[!h]
	
	\centering
	\includegraphics[width=1.00\textwidth]{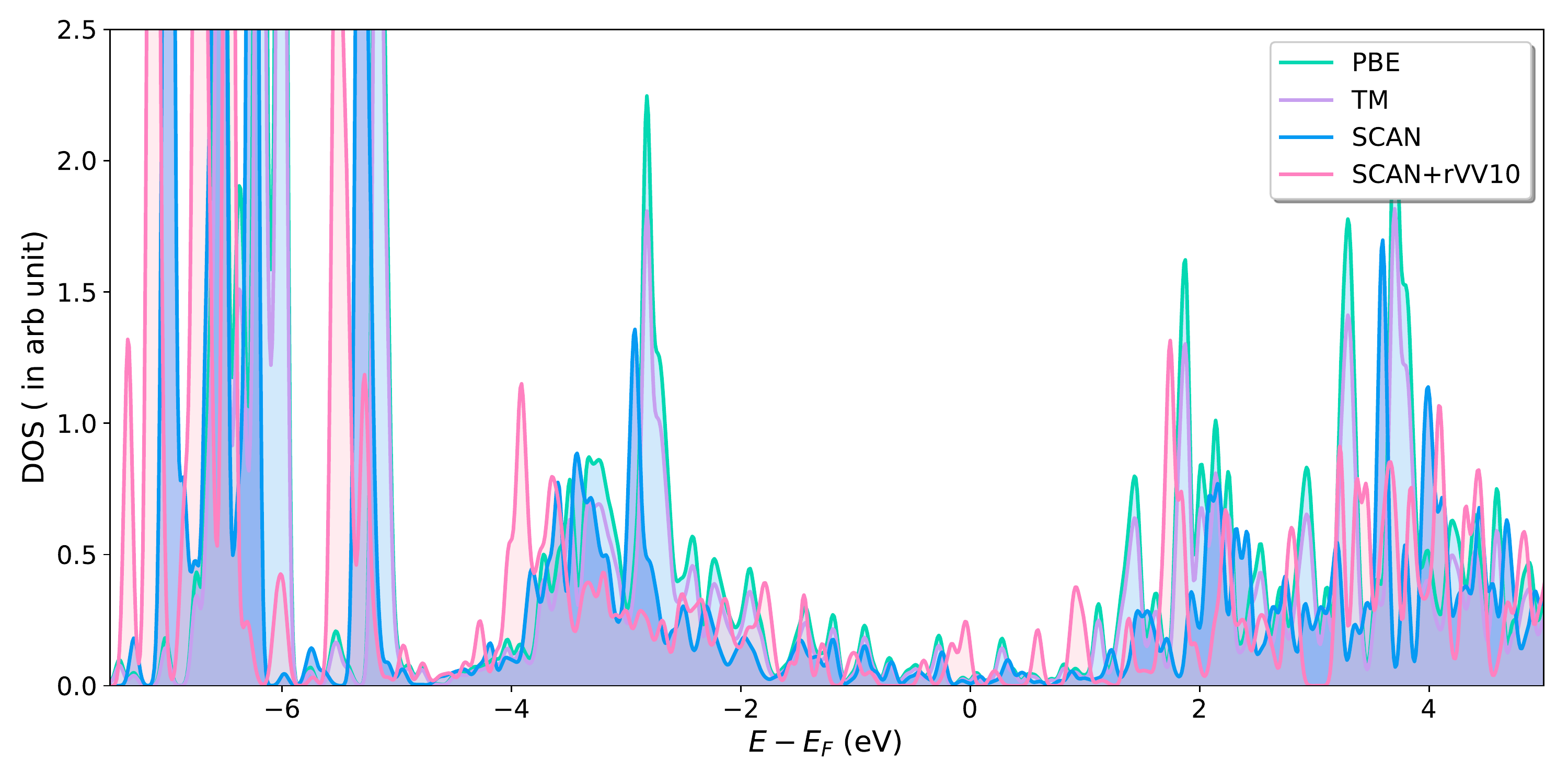}

	\caption{This figure shows the density of plots for adsorbed thiophene molecule on Cu(110) surface calculated using four different density functional approximations. }
	\label{fig:dos-110}
	
\end{figure*}

\FloatBarrier

\FloatBarrier
	\vspace*{0.25cm}
	\section{CONCLUSIONS}
	\vspace*{0.25cm}

	The nature of the adsorption of thiophene on three different copper surfaces was studied. Thiophene is found to adsorb on the top position with the S of the molecule sitting directly above the surface Cu atom. This ``top" configuration has also considered as the energetically favorable geometry in the previously reported theoretical and experimental study. DFT-PBE calculations results presented in this work also agree very well with similar work \cite{callsenthiophene,thiopheneCu110PRLsony,erinjohnsonthiophene}. We also found that the too-small adsorption energies and large metal-molecule distance predicted by PBE are merely an effect of missing long-range vdW interactions. The systematic PDOS and work function study of the three copper surfaces using PBE functional suggests that the degree of hybridization between the molecule and copper surface is very weak causing a slight decrement in metallic work function. This picture completely changes when we used SCAN meta-GGA functional. More accurate exchange and the intermediate-range vdW interactions of SCAN can partially capture the long-range nature of the attractive vdW interactions. Therefore, we can see the adsorption energies predicted by SCAN are in the weak chemisorption regime and good agreement with the experimental finding. Although SCAN is not a true vdW corrected functional, an unexpected error cancellation between its exchange and correlation can give right adsorption energy and binding distance mimicking the behavior of vdW functional like PBE-D2 \cite{callsenthiophene}. The shorter binding distance from SCAN calculation causes a broadening in molecular orbitals and a shift towards more negative energy. For the same reason, the change in work function due to the adsorption-induced dipole at the interface predicted by SCAN is slightly larger compared to PBE. The chemisorption picture becomes clearer when a non-local vdW correction is added self-consistently as we can see in our SCAN+rVV10 results. SCAN+rVV10 overestimates the adsorption energies by 20\% for Cu(111) and by 24\% for Cu(110) and 14\% in case of Cu(111) surface. This tells us that the strength of chemical bonding varies for different electronic arrangement e.g., in a hierarchical order of Cu(100) $>$ Cu(110)  $>$ Cu(111). Furthermore, a large downward shift in the molecular orbitals of SCAN+rVV10 plots in Fig. \ref{fig:dos-110} further facilitates the fact that vdW interactions at the interface can change the electronic arrangement of the metallic surface in the presence of an acceptor or donor molecule. It is important to note that both SCAN and SCAN+rVV10 perform poorly in the much more studied and complex CO adsorption problem \cite{co-patra}. We understand that the vdW interactions play an important role in thiophene-copper bonding compared to the phenomenological charge-transfer problem of CO adsorption on metals. We further explored the performance of the TM meta-GGA functional for this adsorption problem. The exchange of TM meta-GGA is developed using a density matrix expansion of the exchange hole while it's correlation is a modified version of TPSS meta-GGA. This special set-up made TM a very accurate functional to predict jellium surface energies \cite{MO201738}. However, our study showed that TM does not perform very well describing the weak chemisorption nature of thiophene bonding with the copper surface. It has recently been shown that vdW interactions acting at the interface of graphene and metal is very non-local nature \cite{PATRA-TAO, Tang2019} can be well described within the pairwise scheme. We suggest adding such a vdW correction term to TM functional may lead to a better description for adsorption problem.

	\vspace*{0.3cm}
	\section{ACKNOWLEDGMENTS} 
	\label{sec:ackno}
	\vspace*{0.3cm}
	This work was supported by the U.S. National Science Foundation under Grant No. DMR-1607868 (CMMT – Division of Materials Research, with a contribution from CTMC – Division of Chemistry).
	
	\vspace{0.5cm}

	\FloatBarrier	
	\bibliography{thesis-ref}{}
	\bibliographystyle{unsrtnat}
	
\end{document}